\documentstyle[preprint,aps]{revtex}
\begin{document}
\draft
\preprint{\vbox{
\hbox{IFT-P.075/98}
\hbox{hep-ph/yymmddd}
\hbox{October 1998}
}}

\title{ Dark matter:  the top of the iceberg?\thanks{Contributions to Topics 
in Theoretical Physics: Festschrift for A. H. Zimerman  }} 
\author{G.~E.~A.~Matsas, J.~C.~Montero, V.~Pleitez and D.~A.~T.~Vanzella}

\address{ 
Instituto de F\'\i sica Te\'orica\\ 
Universidade Estadual Paulista--UNESP\\ 
Rua Pamplona 145\\
01405-900 -- S\~ao Paulo, SP, Brazil}

\date{\today} 
\maketitle 

\begin{abstract} 

We argue in favor of the existence of the LIPs, 
{\em the least interacting particles}, which would only 
interact with ordinary matter through gravitational field
and could account for (at least) part of the dark matter. 
The de\-tec\-ta\-bi\-li\-ty of LIP matter is addressed at the end.

\end{abstract} 

\newpage

The last fifty years have witnessed continuous efforts towards
the solution of what can be fairly considered the most important
puzzle in Cosmology, namely,  most matter of the Universe
resists to show up except through its gravitational effects.
This is the so--called dark--matter problem.  In fact, as we
scrutinize the Universe at larger and larger scales, we conclude
that 95\% (or even more) of its total mass is hidden from direct
observation.

This missing amount of matter could be expected under
theoretical grounds as follows:  present observations show that
the energy density $\rho$ in the form of luminous matter is about
1\% of $\rho_c$, the critical energy density just necessary to
close the Universe.  Now, if most matter in the Universe was in
this form, and we traced back in time $\rho( t )$, we would find
that near the Big Bang the energy density of the Universe would
be {\em extremely} close to the critical value.  Such a fine tuning
seems unnatural, and it is a current prejudice to assume that
$\rho = \rho_c$.  In such a case, the Universe would be flat
along its whole history, and an enormous amount of non--luminous
matter should presently exist  in the Universe in order to
account for the missing mass\cite{peebles}.

The fact that we would be living today in a Universe with
energy density {\em extremely} close to the critical value is also
predicted by Inflation.  The inflationary theory  became
very  popular recently because of giving a natural explanation
for several cosmological problems -- like the horizon and
monopole problems -- and for providing a nice explanation for the
inhomogeneities observed by COBE in the primordial cosmic
radiation\cite{cobe}.  Notwithstanding, some points are still open, as for
example, the necessary scalar field to drive inflation\cite{inflation} finds no
room in the elementary particle standard model. 

It is known that some dark matter should be in baryonic form, but
this cannot account for all of it.  The energy density due to
baryons cannot exceed about 10\% of $\rho_c$ in order that
nucleosynthesis calculations in the primordial fireball fit
the current abundance of light elements.  So, according to
measurements taken over length scales above 1 Mpc, there must
exist at least as much non--baryonic dark matter as baryonic
matter.  But what could non--baryonic dark matter be?

Maybe the most widespread conjecture is provided by
supersymmetry, which predicts a number of weakly
interacting massive particles -- WIMPs\cite{susy}. 
Attempts to detect these particles are being made without
conclusive results up to now.  Many experiments are currently
under investigation, and more precise results are being
expected, but how to proceed if all  attempts to detect
WIMPs eventually fail? 

We would like to suggest the possibility that the failure in
detecting WIMPs, and other dark-matter candidates\cite{axions} at Earth
laboratories might indicate that non-baryonic dark matter could
be constituted of particles that {\em do not} interact with
ordinary matter by any means, {\em except} gravitationally.  Let
us call them the ``least interacting particles'' (LIPs).  
Although exotic at a first glance, particular realizations of this 
conjecture have already appeared in the literature  
-- like some kinds of {\em mirror matter}, introduced to restore the 
parity symmetry \cite{mp}, and the {\em shadow matter} 
which comes from the superstring--inspired $E^8 \times
E^{8^\prime}$ effective gauge  theory\cite{e8}, but in these cases 
the LIP universe would be restricted to be a perfect copy of our Universe. 
Here, we
would like to argue  in favor of the LIPs, and possible ways of their
detection, in a
model--independent manner in order not to contaminate the discussion
with bias which in general are more related with the underlying 
theories which give rise to LIP matter than with the 
``LIP conjecture'' itself. Indeed, we aim to show that the existence of the LIPs
would be a quite natural and attractive possibility in its own right.

In order to introduce the LIPs in a model--independent framework, 
let us assume that after all symmetry breakings -- which are
expected to
happen in the very first second -- of an
yet unknown all-embracing grand unification 
theory, the relevant world action would be 
$$
S=\int d^4x\sqrt{-g}({\cal R}+{\cal
L}_0+{\cal L}_{\mbox{\small LIP}}), 
$$
where
$g=det(g_{\mu\nu})$,
${\cal R}$ is the curvature scalar,
${\cal L}_0$ is the ordinary--matter Lagrangian density, 
and ${\cal L}_{\mbox{\small LIP}}$ is
some Lagrangian density associated with the  LIPs.  The only
restrictions that ${\cal L}_0$ and ${\cal L}_{\mbox{\small
LIP}}$ must satisfy is that they are scalar functions under
general coordinate transformations, they do not have any fields
in common, and  ${\cal L}_0$ reduces to the usual standard model
in flat spacetimes. It is clear from the action that the
coupling between LIP and ordinary matter is indirect through the
metric, {\em i.e.} they only interact
gravitationally through the spacetime curvature. The spacetime
curvature is determined by the energy content associated with
the fields as  ruled by Einstein  equations.

Now we address the
problem of how likely is the detection of LIP matter.  
It is very difficult to observe the LIPs on Earth--based laboratory
experiments. It is clear from the action above that 
the annihilation of an ordinary--matter pair into a LIP pair through  
graviton exchange could be observable by looking for missing--energy 
events. However, since gravitons couple to any kind of matter very faintly 
and with the same strength (because of the equivalence principle), this kind 
of process would not be observed in practice. 
Thus, it is desirable  that we discuss the detectability issue  
through cosmological and astrophysical observations.

Under some assumptions it is possible  to fix constraints 
on LIP matter by using primordial 
nucleosynthesis and baryogenesis\cite{kt}. This is because
LIP matter would increase the degrees of freedom of the primordial
plasma modifying (i) the energy content of the Universe and thus (ii) 
its expansion rate which is crucial for nucleosynthesis and baryogenesis.

The extra degrees of freedom introduced in form of
LIP matter would be relevant not only cosmologically but also
astrophysically.
Hawking has shown  that black holes must radiate particles in a
thermal spectrum\cite{sh} and might completely evaporate.  A static black
hole of mass $M$ has an area $A= 16\pi M^2 $, and a temperature
$ T =1/ 8\pi M $ as  measured by asymptotic observers. 
In good approximation, a black hole radiates as a
black body.  Using the Stephan-Boltzman law, $dE/dt = a(M) A T^4$,
for the black hole, we obtain that the black--hole mass evaporates 
as  
$$
\frac{dM}{dt} = - \frac{ a(M) }{M^2} ,
$$ 
where $a(M)$ accounts for the degrees of freedom of the emitted
particles.  The more particles a black hole can radiate the
faster the black hole will evaporate.  Thus, if we observed a 
black hole to evaporate faster than what is
predicted by using the standard model of particles, 
it could be indicating that part of the emitted
energy is in LIP form.  (Because of the no-hair
theorems, a black hole would evaporate by the emission of
LIP matter as well as ordinary matter, no mind whether
it was originally formed by the collapse of ordinary or LIP matter.)

A nicer scenario would be provided if 
${\cal L}_{\mbox{\small LIP}}$ gave rise to structures, 
and eventually solved the dark--matter problem. 
Assuming the hypothesis that most mass of the Universe hidden as
dark matter is in  LIP form, and since the most important
interaction in large--scale for structure formation is gravity, we
should expect that most of the inhomogeneity of the observable ordinary
matter would be determined by the inhomogeneity of  LIP matter itself.  
In this vein, it would be natural to expect that large--scale ordinary--matter
structures like observable galaxies, galaxy clusters, {\em etc},
would be located in regions with high LIP--matter concentrations.
Indeed, roughly speaking, the larger the scale the more dark matter is
gravitationally ``observed''. It is possible
to speculate in this context that the so--called 
great--attractor could be constituted 
of  LIP matter.  Clearly, short--scale structures,
like solar systems are formed by local processes like
supernovae, and because of the low density of the Universe the
probability of finding LIP-- and ordinary--matter structures 
overlapping at short scales is very small. 

Furthermore
LIP stellar objects could be observed through the 
lensing effect. 
Typically, gravitational lensing events are associated with extra--galactic 
sources where, for example, the radiation emitted by quasars is bent
by the gravitational field of a foreground galaxy. For our purposes, however, 
it is useful to consider events associated with the lensing of
galactic compact objects. In fact, the MACHO program\cite{macho} is searching 
for galactic dark matter in the form of massive compact halo objects.
Let us suppose  a
sufficiently massive object, like a neutron star,
lying {\it almost} collinear between
some emitting source and Earth in such a way that a double image of the
emitting source is observed.
If the massive object was made of LIP matter we should also
expect an extra luminous point between those two images because
of the light ray that would pass through it.  Thus the detection of 
three clone images in the sky with same $z$ via lensing effect
of a galactic massive object would be a very stimulating 
feature for the  search of LIP matter.

Up to here, we have suggested that the LIPs can be a reasonable 
solution to the dark--matter problem.  Now, we would like to argue 
in favor of their existence in their own right:  Quarks interact through strong,
electromagnetic, weak and gravitational (as far as we know)
interaction; charged leptons, {\em e.g.}  electrons, interact through
electromagnetic, weak and gravitational interaction; 
neutrinos interact through weak and
gravitational interaction.  Why should not we expect to exist a
class of matter particles that would only interact through
gravitational interaction with ordinary matter?  If this is true
that nature follows the motto that everything that is not
forbidden is mandatory, it is hard to believe that at least part
of the energy content of the Universe is not in  LIP form.  If
this reasoning was used fifty years ago to conjecture the
existence of the LIPs, dark matter would be its most outstanding
prediction.  Moreover the LIPs can give us clues for the solution of
other problems. For example, the LIP sector of the Lagrangian 
provides natural room for particles like scalars which not only
are important for Inflation (to cite just an example), but also
are expected to exist under ``naturality'' grounds.

In summary, we think that even if experiments  rule out the
existence of the LIPs, the fact that Nature has chosen not to
realize particles that interact only gravitationally must be
indicating some very deep fact. Using Bohr's words: {\em ``The
opposite of a deep statement is another deep statement''}.
It seems to us that it is more natural to hold that the LIPs exist,  
than the opposite.

\section*{Acknowledgments}

This research was supported in part by the Conselho Nacional
de Desenvolvimento Cient\'{\i}fico e Tecnol\'ogico (GM, JM
and VP), and in part by Funda\c{c}\~ao de Amparo \`a Pesquisa
do Estado de S\~ao Paulo (DV).

\end{document}